\documentclass[conference]{IEEEtran}
\IEEEoverridecommandlockouts
\usepackage{cite}
\usepackage{amsmath,amssymb,amsfonts}
\usepackage{graphicx}
\usepackage{textcomp}
\usepackage{xcolor}

\usepackage{mathptmx} 

\newcommand{\ignore}[1]{}
\usepackage{fancyhdr}
\usepackage[normalem]{ulem}
\usepackage[hyphens]{url}

\usepackage{microtype}

\usepackage{siunitx}

\usepackage{booktabs} 
\usepackage[boxruled,linesnumbered,noend]{algorithm2e}

\usepackage{comment}
\usepackage{pifont}
\usepackage{multirow}

\usepackage{floatrow}
\usepackage{setspace}
\usepackage[export]{adjustbox}

\def\BibTeX{{\rm B\kern-.05em{\sc i\kern-.025em b}\kern-.08em
    T\kern-.1667em\lower.7ex\hbox{E}\kern-.125emX}}
    
\begin{document}
\title{Shenjing: A low power reconfigurable neuromorphic accelerator with partial-sum and spike networks-on-chip\thanks{This research is supported in part by Programmatic grant no. A1687b0033 from the Singapore government's Research, Innovation and Enterprise 2020 plan (Advanced Manufacturing and Engineering domain).}\thanks{
This work is accepted by Design, Automation, and Test in Europe (DATE 2020). Grenoble, France. Mar 2020.}} 
\author{Bo Wang\textsuperscript{*}, Jun Zhou\textsuperscript{*}\thanks{*Both authors contributed equally to this work.}, Weng-Fai Wong, and Li-Shiuan Peh\\
        School of Computing, National University of Singapore}

\maketitle
\begin{abstract}
The next wave of on-device AI will likely require energy-efficient deep neural networks. Brain-inspired spiking neural networks (SNN) has been identified to be a promising candidate. Doing away with the need for multipliers significantly reduces energy. 
For on-device applications, besides computation, communication also incurs a significant amount of energy and time. 
In this paper, we propose Shenjing, a configurable SNN architecture which fully exposes all on-chip communications to software, enabling software mapping of SNN
models with high accuracy at low power. 
Unlike prior SNN architectures like TrueNorth, Shenjing does not require any model modification and retraining for the mapping.
We show that conventional artificial neural networks (ANN) such as multilayer perceptron, convolutional neural networks, as well as the latest residual neural networks can be mapped successfully onto Shenjing, realizing ANNs with SNN's energy efficiency. For the MNIST inference problem using a multilayer perceptron, we were able to achieve an accuracy of 96\% while consuming just 1.26~mW using 10 Shenjing cores. 
\end{abstract}

%
%
\section{Introduction}
\label{sec:intro}
The phenomenal success of deep learning has led to the use of specialized hardware accelerators such as Google's TPU in data centers, driving cloud-based AI. On-device AI will be next: Applications with stringent timing, form-factor, energy, privacy requirements will demand AI on embedded devices. Compared with ANNs, SNNs are inherently more energy-efficient. First, data passing through the layers are binary, so computation of weighted sum is done through additions, instead of multiplications. This reduces computation power and time significantly. Moreover, a SNN will perform an addition only when the input is $1$, which in practice is very sparse, whereas ANN performs multiplications on all neurons across all inputs. While the compute requirement is ameliorated, the on-chip movement of data, i.e., communication, now dominates cost. 

We propose Shenjing, a reconfigurable SNN accelerator, whose essence lies in its per-neuron software-defined on-chip networks (NoCs) that allow software to configure and link up individual hardware neurons to match myriad neural networks. Summation of weights is efficiently performed enroute within Shenjing's partial sum NoCs, followed by the generation of spikes which are configured to route each neuron to its next layer neuron through Shenjing's spike NoCs. By exposing all on-chip communications to software configuration, 
Shenjing's NoCs can be ultra-lightweight, requiring no buffer queues or virtual channels and no routing or flow control logic. 

Shenjing's software-defined NoCs also enable run-time reconfiguration. While energy-efficient SNNs are particularly suited for on-device AI, they are not currently widely used. Deploying SNNs will require the development and training of new models from scratch, a major hurdle to deployment. Shenjing's reconfigurability allows for the mapping and transfer learning of widely used ANNs onto Shenjing's SNN hardware, bootstrapping SNN deployment.

%
%

%
%

\label{sec:arch}


\begin{figure}[t]
\includegraphics[width=0.9\linewidth]{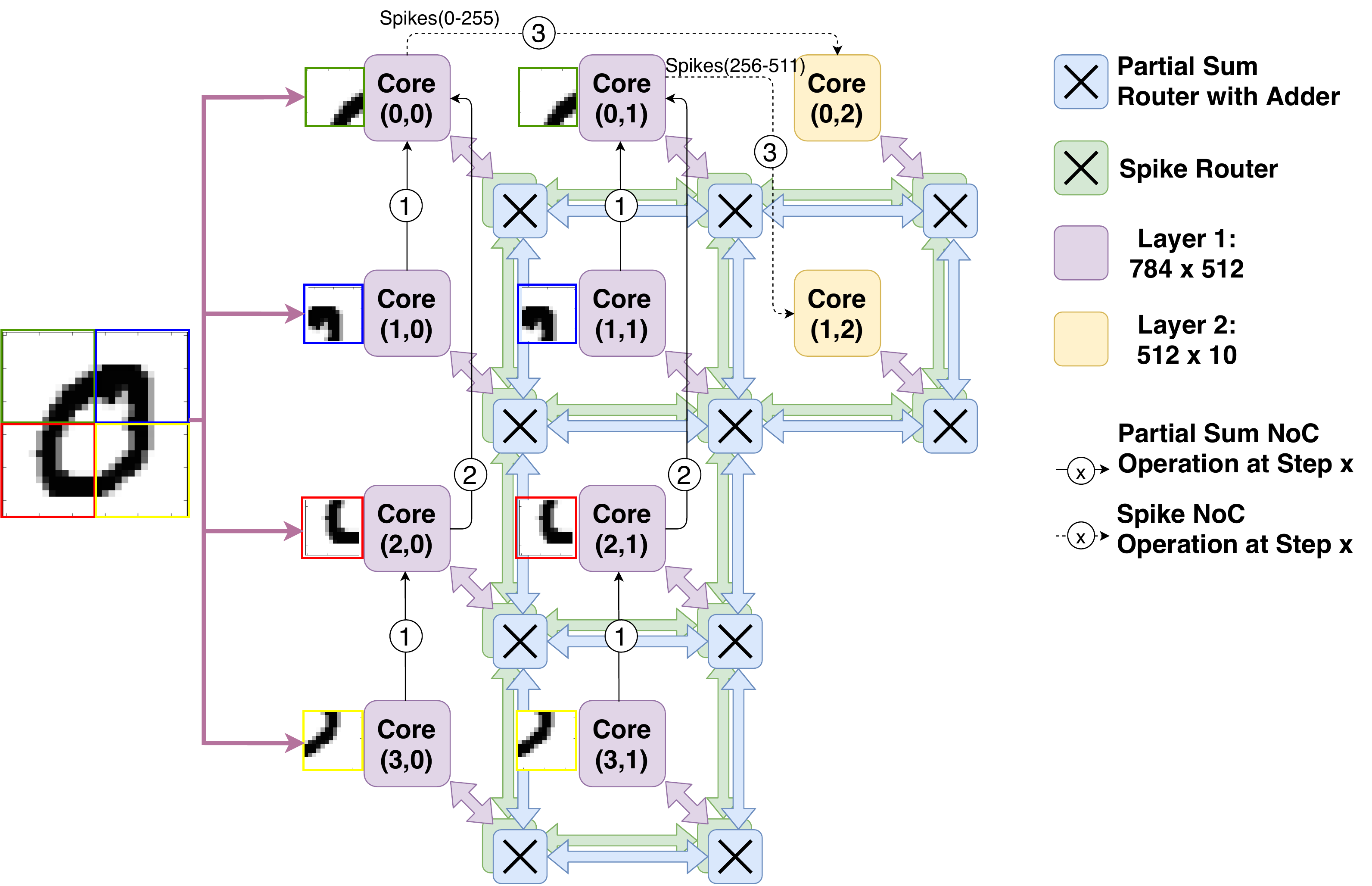}
\caption{Mapping of MNIST-MLP onto Shenjing.}
\label{fig:neuron_core_abstract}
\end{figure}

\section{Architectural overview}
\label{sec:architecture}

In a SNN, two key operations are performed at each layer of neurons: (1) weights of incoming synapses that spiked are summed, and (2) if this sum exceeds a threshold, a spike is sent to the next layer.
Figure~\ref{fig:neuron_core_abstract} illustrates a Shenjing chip with 10 cores arranged in a grid where each core has 256 neurons interconnected by per-neuron NoCs that are meshes. 
Weight summation, is handled first by the neuron core and then via the 256 {\it partial sum NoCs} (PS NoCs) that perform in-network additions across cores. 
Spiking is handled by the 256 {\it spike NoCs} that generate and communicate 1-bit spikes across cores. 

Shenjing's NoCs are software-defined, having the twin benefits of reconfigurability and low hardware overhead. The software mapping tool (see Section~\ref{sec:map}) maps neurons to cores, sets up the adder tree topology, configures the PS NoCs, and connects up the cores with the spike NoCs. The per-neuron NoCs enable simple mapping of neural network models onto Shenjing. 
Software-defined configurations are stored in Shenjing's configuration memories, governing the cycle-by-cycle operation of the hardware. The NoC routers are thus freed from handling any run-time routing or flow control, significantly simplifying their design. 

Figure~\ref{fig:neuron_core_abstract} shows a simple mapping of a fully connected network for MNIST as Table~\ref{table:applications}(a) on Shenjing. 
As each core only holds 256 neurons, the 784 MNIST inputs can be evenly mapped onto 4 cores and the hidden layer of 512 neurons can be mapped onto 2 cores. The spikes from the input layer to the first half of the 512 neurons are mapped onto cores (3,0), (2,0), (1,0) and (0,0). Similarly, the spikes to the other 256 neurons are duplicated and mapped onto cores (3,1), (2,1), (1,1) and (0,1). The output layer fits onto cores (0,2) and (1,2).

First, weighted sum is performed within each neuron core to form the local partial sums. Next, the corresponding local partial sums (PS) from (3,0) and (3,1) are injected into the partial-sum NoCs, to cores (2,0), (2,1) which adds them with their local core's sum, respectively. Cores (1,0) and (1,1) proceed similarly. Finally, at cores (0,0) and (0,1), the PS accumulates with those from cores (2,0) and (2,1) to create the full weighted sum which is fed to the spike generation unit at cores (0,0) and (0,1) to determine if a spike should be fired. 


\begin{figure*}[bth]
\centering
\includegraphics[width=0.9\linewidth]{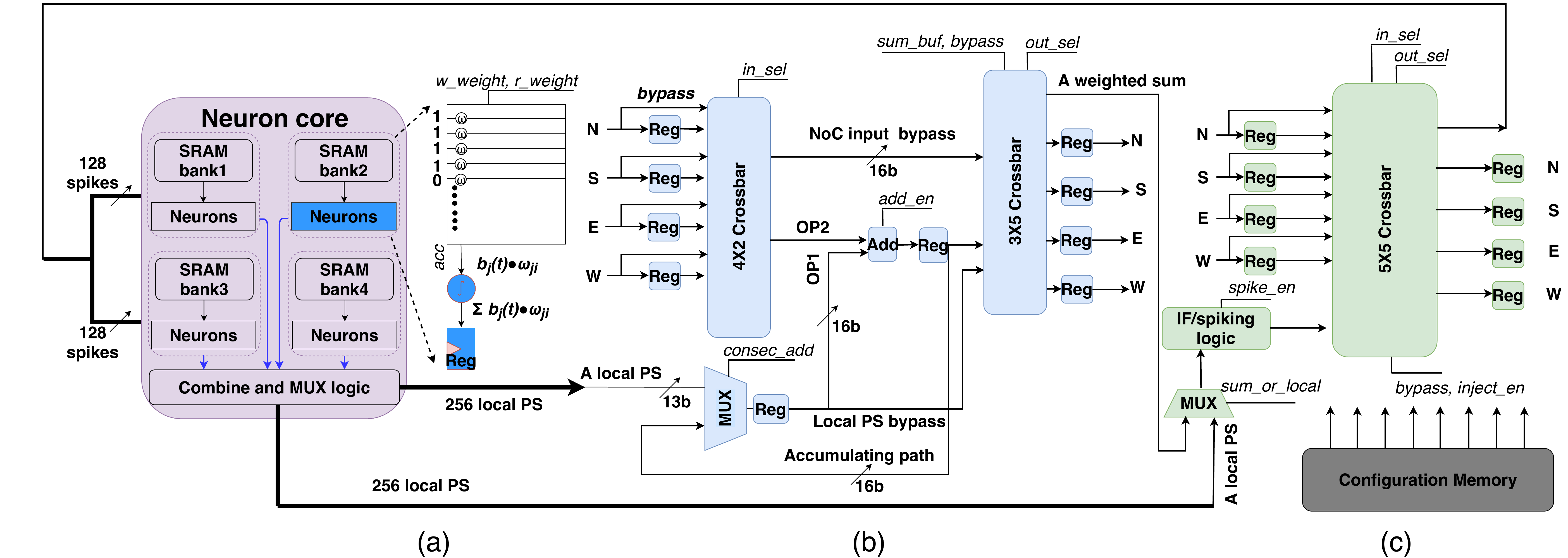}
\caption{Micro architecture of (a) a neuron core (b) a partial sum router and (c) a spike router.}
\label{fig:core_ps_spike}
\end{figure*}

Figure~\ref{fig:core_ps_spike} details the microarchitecture of Shenjing's neuron core, partial-sum and spike routers.

\noindent{\bf Neuron core.}
\label{sec:arch_neuron_core}
Each neuron core is a collection of weight synapses stored in 4 SRAM banks, computing elements (neurons) implemented with accumulators, and control logic~\cite{truenorth}~\cite{Merolla2014TrueNorth}.

\noindent{\bf Partial Sum NoCs.}
A $4 \times 2$ input crossbar fetches input data from a port (North, South, East or West) and either registers it for local addition or bypasses this router to adjacent routers via output links. The $3 \times 5$ output crossbar ejects it to the desired port or the spiking logic of the neuron core. Our mapping software constructs a PS adder tree by configuring the crossbars of the PS NoC router appropriately. These configuration bits are loaded into a configuration memory prior to Shenjing execution, and drives the select signals of the crossbars and muxes. Each PS NoC is dedicated exclusively to the same neuron in each core, and the software mapping tool maps neurons appropriately (see Section~\ref{sec:map}). Hence, unlike conventional NoCs, the PS NoC does not need buffer queues, flow control or routing logic.

PS NoCs' bitwidth corresponds to the data width of the synaptic weight and adders within the routers. When two $n$-bit numbers are added, the sum is $n+1$ bits. Having a 16 bit width allows us to sum up $2^{11}$ 5-bit weights at the worse case where all weights are {\tt 11111} with all input spikes being `1'. In practice, the sparsity of spikes allows even more accumulations. 
We did not encounter any overflow in our applications. 


\noindent{\bf Spike NoCs.} A multiplexer selects either the local partial sum from the neuron core (when a layer fits entirely in a core), or the full weighted-sum arriving from the PS router (when a layer spans multiple cores) as input for the spiking logic unit. The latter integrates the sum and compares the potential against the threshold. If it exceeds the threshold, a spike is generated, and the potential value is subtracted from the threshold. The 1-bit spike then enters the $5 \times 5$ crossbar where it is buffered and sent to the next hop in the spike NoC. 


Shenjing's spike NoCs support hardware multicast. A single spike packet injected by a neuron core can be scheduled by software to be X-Y routed to successive multicast destinations, ejecting the spike when it arrives at each destination in turn. 
Hardware multicast is particularly useful when mapping CNNs onto Shenjing, when spikes from a core are often sent to multiple cores located next to each other. 

 
\noindent{\bf Reconfigurability and accuracy.} 
Prior SNN accelerators~\cite{Merolla2014TrueNorth, spinnaker,shi2015development} chose block-level spike connectivity instead, so an entire block of multiple spikes is routed together to the same destination block. Block-level spike NoCs amortize the hardware cost of dynamic routing and flow control across many spikes. Shenjing is able to support neuron-level bit-wise spike NoCs because our spike NoCs are completely software-defined, so such hardware overheads are not an issue. As most ANNs demand arbitrary connections to cores at diverse locations, Shenjing's per-neuron NoCs provide the most flexible substrate for configuring ANNs onto our SNN hardware.

Previous SNN architectures communicate strictly using spikes across cores, and do not support partial sums. When a layer cannot fit within a core, each core computes a partial sum based on the subset of axons and synapses within the core, then integrate and fire a spike. An aggregating core sums these spikes to gain a representation of full weighted-sum and generates a final output for the layer. This can lead to significant accuracy loss. Prior SNN architectures side-step this by requiring SNN models to be developed and trained from scratch, taking into account hardware core size constraints (see Section~\ref{sec:related-works}). Shenjing eliminates such accuracy loss with its PS NoCs that precisely add weights between cores within the network, thereby allowing software to easily map pre-trained models onto Shenjing hardware.

%
%
\section{Mapping SNN to Shenjing}
\label{sec:map}

\begin{figure}[ht]
\includegraphics[width=0.8\linewidth]{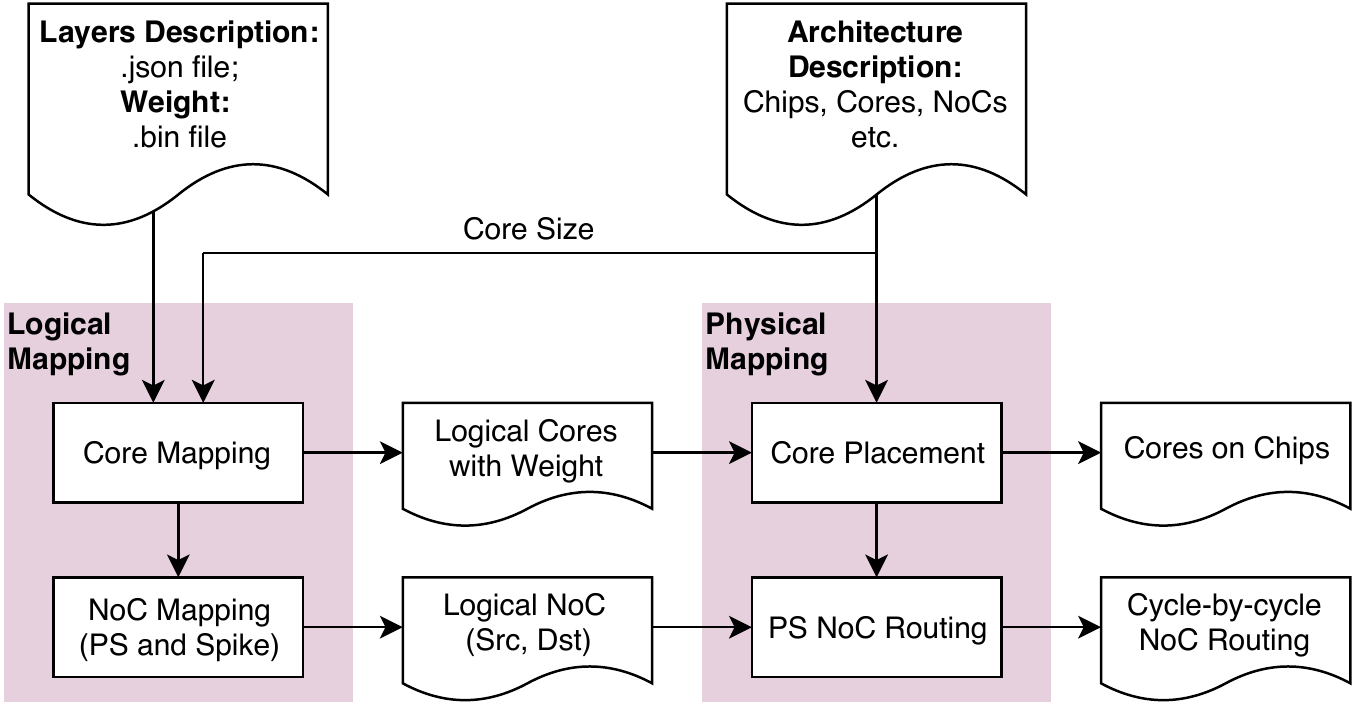}
\caption{Shenjing's software mapping tool flow.}
\label{fig:compiler_overview}
\end{figure}

Figure~\ref{fig:compiler_overview} shows our
toolchain that automatically maps a trained SNN onto Shenjing. In the first logical mapping phase, the toolchain maps each layer's weights onto a set of logical cores, and then schedule the logical PS NoCs to produce the total weighted sum within the layer. It also schedules the logical spike NoCs to direct the spikes among layers. In this phase, NoC scheduling determines just the source and destination. Next, in the physical mapping phase, logical cores are placed onto physical chips, and the logical NoC scheduling is implemented as cycle-by-cycle routing instructions.

\noindent{\bf Logical core and partial-sum NoC mapping.}
When the number of inputs to a layer exceeds the size of a core, we split them into multiple cores, and accumulate the partial weighted sum using the partial-sum NoC. 

1. {\em Mapping fully connected layers}:
To map an $m\times n$ fully connected layer with $m$ inputs and $n$ outputs where both $m$ and $n$ exceed the core size, we need $n_{\mbox{\scriptsize row}} \times n_{\mbox{\scriptsize col}}$ cores, where $n_{\mbox{\scriptsize row}}=\lceil m/N_{\mbox{\scriptsize in}}\rceil$, $n_{\mbox{\scriptsize col}} = \lceil n/N_{\mbox{\scriptsize out}}\rceil$ and $N_{\mbox{\scriptsize in}}$ and $N_{\mbox{\scriptsize out}}$ are the number of synapses and neurons of one core. We arrange the $n_{\mbox{\scriptsize row}} \times n_{\mbox{\scriptsize col}}$ cores in a rectangle. The rows receive the $m$ inputs and columns produce $n$ outputs.
Algorithm~\ref{alg:scheduleFC} shows the logical scheduling of partial-sum NoCs to produce the total weighted sum.
\begin{algorithm}[ht]
\footnotesize
  \caption{Software mapping algorithm for a fully connected layer onto Shenjing's partial sum NoCs. }    \label{alg:scheduleFC}
  \SetKwInOut{Input}{Input}
  \SetKwInOut{Output}{Output}
  \Input{$n_{\mbox{\scriptsize row}} \times n_{\mbox{\scriptsize col}}$ cores in rectangle with local partial-sum PS$(i, j)$.}
  \Output{Network trace $\mathbf{N}$}
  $\mathbf{N} \leftarrow \emptyset$\;
  $\mbox{fold} = \lceil\log_2^{n_{\mbox{\scriptsize row}}}\rceil$\;
  \For{($f\leftarrow 1; f < 2^{\mbox{\normalfont \scriptsize fold}}; f=f*2)$}
  {
  	$\mathbf{L}\leftarrow \emptyset$\;
    \For{($i=f; i<n_{\mbox{\normalfont\scriptsize row}}; i=i+2*\mbox{\normalfont fold} $)}{
    	\For{$j=0, \ldots, n_{\mbox{\scriptsize \normalfont col}} $}{
            	$\mathbf{L}$.add(``Send PS$(i, j)$ FROM $(i, j)$ TO $(i-\mbox{fold}, j)$'')
        }
    }
    $\mathbf{N}$.add($\mathbf{L}$)\;
    $\mathbf{L}\leftarrow \emptyset$\;
    \For{($i=f; i<n_{\mbox{\normalfont\scriptsize row}}; i=i+2*\mbox{\normalfont fold} $)}{
    	\For{$j=0, \ldots, n_{\mbox{\scriptsize \normalfont col}} $}{
            	$\mathbf{L}$.add(``Add PS$(i, j)$ TO PS$(i-\mbox{fold}, j)$'')
        }
    }
    $\mathbf{N}$.add($\mathbf{L}$).
  }
  
  \Return $\mathbf{N}$\;
\end{algorithm}

2. {\em Mapping convolution layers}:
For a convolution layer with input size $h\times w \times c_{\mbox{\scriptsize in}}$ and kernel size $k\times k \times c_{\mbox{\scriptsize in}}\times c_{\mbox{\scriptsize out}}$, we need $n_h \cdot n_w$ cores to cover the spatial dimension $h\times w$ where $n_h = \lceil h/(\sqrt{N_{\mbox{\scriptsize in}}}-2(k-1))\rceil$ and $n_w = \lceil w/(\sqrt{N_{\mbox{\scriptsize in}}}-2(k-1))\rceil$. So, to cover all input and output channels, a total of $c_{\mbox{\scriptsize in}}\cdot c_{\mbox{\scriptsize out}}\cdot n_h \cdot n_w$ cores is needed. The weights are mapped so that each neuron performs the convolution in one channel as its partial sum.

To produce the convolution sum, along each channel, the neighbouring cores first exchange their partial-sums to produce the convolution of the boundary pixels. Then, among the channels, the partial sums are accumulated to complete the convolution.
As a convolution layer scans through the input image, there will be overlap at the boundary of each core.

\begin{figure}[t]
\includegraphics[width=0.65\linewidth]{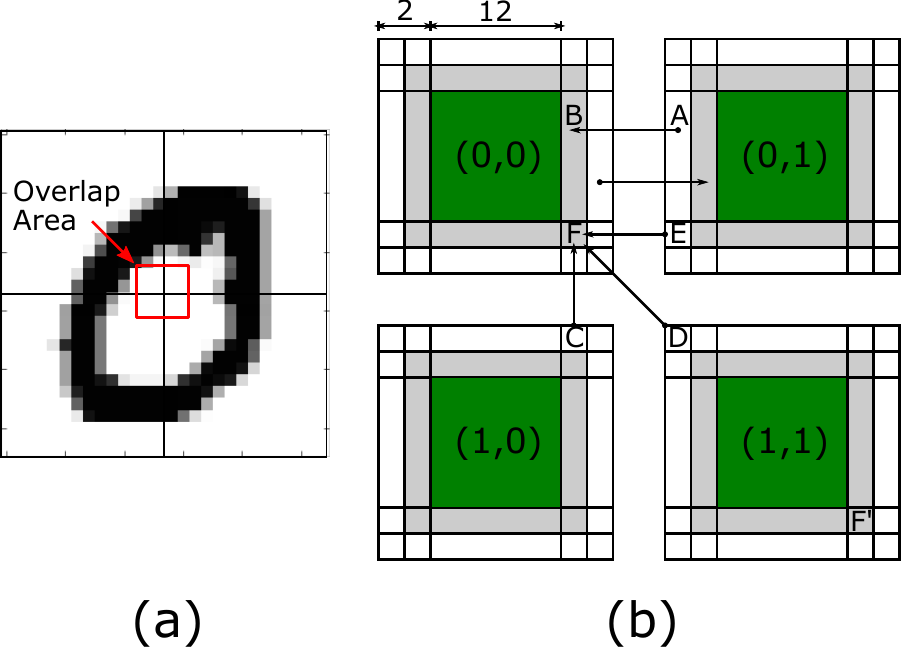}
\caption{Mapping of a convolution layer, (a) viewed in 2D data and (b) mapped onto 4 Shenjing cores.}
\label{fig:conv_map}

\end{figure}

As an example, to map a $3\times 3$ kernel over an MNIST image to cores of $256$ synapses and $256$ neurons (Figure~\ref{fig:conv_map}), we split the $28\times28$ input into four $14\times 14$ rectangles and allocate each quadrant to a core, as shown in (a). Computing the output in the overlapping corners of each core, highlighted in a red box, requires input from all four cores. These overlapped data has to be duplicated and supplied to each of the four. In this example,
each core produces $12\times 12$ complete weighted sums (green areas), slices of $2\times 12$ partial sums at each of the four boundaries and $2\times 2$ partial sums at each of the four corners. At the boundary of core (0,0) and (0,1), sending (0,1)'s outer partial sum~(A) to (0,0) and adding to its inner partial sum~(B) will produce the complete sum with size $1\times 12$. At the corner, sending the partial sums~(C, D, E) to (0,0) and adding to F will produce the complete sum of size $1\times 1$. The operations are symmetric among the neighboring cores. As a result, each core produces $14\times 14$ outputs. 

The partial sums in Figure~\ref{fig:conv_map} (A-F) are routed by the partial sum NoCs, where the data to be accumulated must be produced by the neurons in the same channel, e.g. areas (C-F) must be produced by neuron \#1 of all four cores. Consider F' area in core (1,1). Since area D has occupied neuron \#1, F' has to be allocated to a different neuron from F. This requirement applies to all partial sum areas and results in an inter-changing pattern of neuron allocation among neighboring cores.

3. {\em Mapping ResNet shortcuts}: Converting ResNet to its spiking version was proposed in~\cite{Hu_spike_resnet} where a layer's output skips its following residual block and directly adds to the block's output as a shortcut. In SNN, the shortcut needs to be normalized so we add a  normalization layer in Shenjing with weights set to a diagonal matrix $\text{diag} (\lambda)$. The partial sum after normalization is then sent to the corresponding cores of the residual block through PS NoCs for addition. Shenjing's PS and spike NoCs naturally support ResNet-style shortcuts between non-adjacent cores, through intermediate routers. 

\noindent{\bf Logical spike NoC mapping.}
In general, the mapping of spike NoCs among layers is straightforward because the output size of cores naturally fit the input size. So, we usually just give a core-to-core mapping. However, when the output size of each core in a layer is small enough, multiple cores' output can be directed to one core in the subsequent layer. For this purpose, we map the output of multiple cores to different non-overlapping neurons so they can be sent to the same core.

\noindent{\bf Physical Mapping.}
For every layer, the logical cores are then mapped to the Shenjing architecture. We implemented a greedy algorithm to allocate adjacent layers next to each other, while attempting to minimize the number of chips (of 784 cores each) used and the cost of data movement. In particular, for each layer, we first search for a rectangular space that can accommodate this layer among existing chips.
If that is not available, we then add Shenjing chips to house this layer.
For the physical routing of the logical NoC schedule, we use simple deterministic XY routing. For flow control, a packet (spike or PS) is scheduled to wait if the output port/link is occupied, as there are no buffer queues. After physical mapping, a cycle-by-cycle compiled schedule of atomic operations onto binary signals that control Shenjing's hardware components (see Table~\ref{table:op2sig}) is generated.    

\begin{table}[t]
\centering
\footnotesize
\vspace{-1em}
\begin{tabular}{|l|l|}
\hline
\multirow{2}{*}{\textbf{Partial Sum Router}} & \textbf{type{[}2{]}}  \textbf{sum\_buf}   \textbf{add\_en} \\            
& \textbf{consec\_add}  \textbf{bypass}        \textbf{in\_sel{[}2{]}}  \textbf{out\_sel{[}3{]}} \\ \hline
{\tt SUM \$SRC, \$CONSEC}&            {\tt 00}                      {\tt 0}                   {\tt 1}                            {\tt \$CONSEC}              {\tt 0}                      {\tt \$SRC}                     {\tt 000}                        \\ \hline
{\tt SEND \$SRC, \$DST}            & {\tt 00}                      {\tt \$SRC}               {\tt 0}                            {\tt 0}                     {\tt 0}                      {\tt 00}                         {\tt \$DST}                      \\ \hline
{\tt BYPASS \$SRC, \$DST}          & {\tt 00}                      {\tt 0}                   {\tt 0}                            {\tt 0}                     {\tt 1}                      {\tt \$SRC}                      {\tt \$DST}                      \\ \hline\hline
\multirow{2}{*}{\textbf{Spike Router}}       & \textbf{type{[}2{]}}  \textbf{spike\_en}  \textbf{sum\_or\_local}     \\ 
& \textbf{inject\_en}   \textbf{bypass}        \textbf{in\_sel{[}2{]}}  \textbf{out\_sel{[}2{]}} \\ \hline
{\tt SPIKE \$SUM\_OR\_LOCAL}                & {\tt 01}                      {\tt 1}                   {\tt \$SUM\_OR\_LOCAL}                        {\tt 0}                     {\tt 0}                      {\tt 00}                         {\tt 00}                         \\ \hline
{\tt SEND \$DST}                 & {\tt 01}                      {\tt 0}                   {\tt 0}                            {\tt 1}                     {\tt 0}                      {\tt 00}                         {\tt \$DST}                      \\ \hline
{\tt BYPASS \$SRC, \$DST}          & {\tt 01}                      {\tt 0}                   {\tt 0}                            {\tt 0}                     {\tt 1}                      {\tt \$SRC}                      {\tt \$DST}                      \\ \hline\hline
\multirow{2}{*}{\textbf{Neuron Core}}             & \textbf{type{[}2{]}}  \textbf{r\_weight}  \textbf{w\_weight{[}4{]}}    \\   
& \textbf{acc{[}4{]}}      \textbf{pad{[}5{]}}                          \\ \hline
{\tt LD\_WT}                       & {\tt 10}                      {\tt 0}                   {\tt 1111}                                         {\tt 0000}                {\tt 00000}             \\ \hline
{\tt ACC}                         & {\tt 10}      {\tt 1}                                 {\tt 0000}                                        {\tt 1111}            {\tt 00000}                     \\ \hline
\end{tabular}

\caption{Mapping of atomic operation to hardware control signals. \newline
{\normalfont There are three types of atomic operations for partial sum router, spike router and neuron core, defined by the first 2 bits. {\tt \$SRC} and {\tt \$DST} are one of {\tt N}(orth), {\tt S}(outh), {\tt E}(ast), or {\tt W}(est). {\tt \$CONSEC} controls the multiplexor in Figure~\ref{fig:core_ps_spike}~(b) that feeds either the local PS value (0) or the previous sum (1) as the first operand into the adder.}}
\label{table:op2sig}
\end{table}

%
%
\section{Synthesis results of Shenjing}
\label{sec:synthesis}

We implemented our design in system Verilog and synthesized it into gate-level netlist on a 28nm process. We focus on MNIST in this section due to RTL simulation tractability; Larger network results are shown in the next section.

\noindent{\bf Area.} Each tile (neuron core + NoC routers) in Shenjing is synthesized at supply voltages of 0.85V for logic gates and 1.05V for SRAMs, respectively, into 0.262 million logic gates with a total cell area of 0.49 mm\textsuperscript{2}. 
We assume a maximum die size of 20mm by 20mm, so 784 Shenjing tiles can be fit onto a chip in a 28-by-28 grid.  
Our routers take up a sizable portion of tile area (39\%), comparable to the SRAMs (44\%) as they perform computations of sum and spikes as well.

\begin{figure}[tb]
\includegraphics[width=0.8\linewidth]{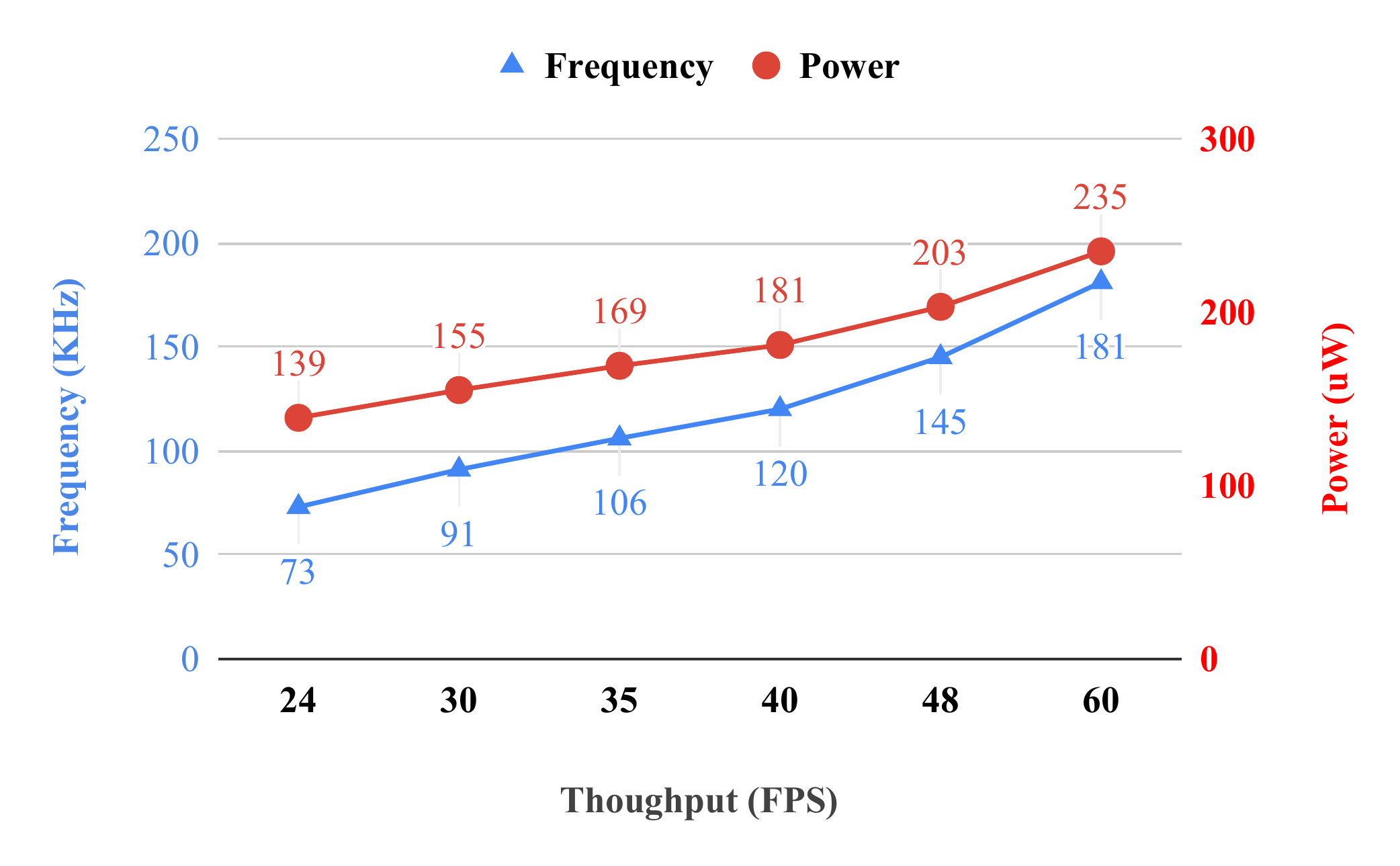}
\vspace{-0.5cm}
\caption{Tradeoff of Shenjing's overall throughput with clock frequency and power of a single tile (neuron core + NoC routers)}
\label{fig:fps_freq_pwr}
\end{figure}

\noindent{\bf Interactions of neural net throughput with circuit timing and power.} The critical path begins from the accumulative registers in the neuron circuit and ends at the pooling registers located in the IF/Spiking logic. 
The maximum achievable frequency of Shenjing is 243MHz. 
Our work is targeted to process the FC and convolutional layers at a minimum throughput of 30 frames/sec (fps) which suffices for typical video applications deploying CNNs. 
Complexity of the SNN model impacts the chip frequency needed to achieve this. Figure~\ref{fig:fps_freq_pwr} shows the tradeoff of throughput with frequency and power of a tile. 
As throughput is increased from the 24 to 60fps, the required operating frequency ($F_{req}$) increases from 73kHz to 181kHz. Power  dissipation scales linearly with $F_{req}$, increasing $2.48 \times$ from 139$\mu$W to 235$\mu$W.
Here we chose a target video throughput of 40fps for MNIST MLP for a balance of power and performance.

\begin{table}[bt]
\footnotesize
\vspace{-1em}

\begin{tabular}{|l|l|S[table-format=1.4]|S[table-format=3.2]|}
\hline
\multicolumn{1}{|c|}{\textbf{Block}}                                                                         & \multicolumn{1}{c|}{\textbf{Atomic Op}} & \textbf{\begin{tabular}[c]{@{}c@{}}Active power\\ @120~kHz \\ (mW)\end{tabular}} & \textbf{\begin{tabular}[c]{@{}c@{}}Active energy\\ per neuron\\ (pJ)\end{tabular}} \\ \hline
\multirow{3}{*}{\textbf{\begin{tabular}[c]{@{}l@{}}Partial sum\\ router\end{tabular}}} & SUM             & 0.0383                                                                          & 1.25                                                                               \\ \cline{2-4} 
                                                                                       & SEND            & 0.0443                                                                          & 1.44                                                                               \\ \cline{2-4} 
                                                                                       & BYPASS          & 0.0455                                                                          & 1.48                                                                               \\ \hline
\multirow{3}{*}{\textbf{Spike router}}                                                 & SPIKE           & 0.0689                                                                          & 2.24                                                                               \\ \cline{2-4} 
                                                                                       & SEND            & 0.0721                                                                          & 2.35                                                                               \\ \cline{2-4} 
                                                                                       & BYPASS          & 0.0381                                                                          & 1.24                                                                               \\ \hline
 \textbf{Neuron core$^1$}                                                                                         & ACC$^2$             & 0.0412                                                                          & 171.67                                                                             \\ \hline
\textbf{Initialization}                                                       & LD\_WT$^2$          & 0.0568                                                                          & 236.67                                                                             \\ \hline
\end{tabular}
\caption{Synthesized active power and energy of atomic operations}

$^1$ SRAM power was obtained at 1.05V while the other logic operates at 0.85V due to the limitations of the SRAM libraries we have access to.\\
$^2$ Active energy of LD\_WT (loading of weights) which occurs only once during initialization, and ACC (accumulation across a subcore) is higher as these operations take 131 cycles while the others take a cycle.
\label{table:energy}
\end{table}

\noindent{\bf Power.} We performed cycle accurate power analysis using Prime Time PX for MNIST MLP running on 10-core Shenjing and the overall power reported by the tool was 1.26mW. 

Table~\ref{table:energy} lists each atomic operation's energy by PrimeTime with the switching activity of MNIST MLP (6.25\%), calculated using average number of spiking axons per core in each time step. 
These RTL power estimates are fed into our system-level functional simulations for power estimation of larger networks.

%
%
\section{System level results}
\label{sec:sys}

We developed a cycle-level functional simulator for simulating large scale neural network benchmarks that are beyond the ability of detailed RTL simulation. It is aimed to be cycle-by-cycle equivalent to RTL simulation in (1)~running input atomic operations as defined in Table~\ref{table:op2sig}, (2)~producing and routing data in neuron cores and NoCs, and (3)~providing execution statistics for deriving architectural power estimates.

\begin{table}[ht]
\centering
\footnotesize
\vspace{-1em}

\begin{tabular}{|c|c|c|c|}
\hline
{(a) MNIST}
& {(b) MNIST} 
&{(c) CIFAR-10}
&{(d) CIFAR-10}\\
{MLP}
& {CNN} 
&{CNN}
&{ResNet} \\ \hline

\hspace{-0.6em}\begin{tabular}[c]{@{}l@{}}Input(28, 28, 1)\\ FC1(784, 512)\\ FC2(512, 10)\end{tabular} \hspace{-0.8em}
&\hspace{-0.6em}\begin{tabular}[c]{@{}l@{}}Input(28, 28, 1)\\ Conv1(3,3,1,16)\\ Pool1(2,2)\\ Conv2(3,3,16,32)\\ Pool2(2,2)\\ FC1(1568, 128)\\ FC2(128, 10)\end{tabular} \hspace{-0.8em}
&\hspace{-0.6em}\begin{tabular}[c]{@{}l@{}}Input(24, 24, 3)\\ Conv1(5,5,1,16)\\ Pool1(2,2)\\ Conv2(5,5,16,32)\\ Pool2(2,2)\\ Conv3(3,3,32,64)\\ Pool3(2,2)\\ FC1 (576, 256)\\ FC2(256, 128)\\ FC3(128, 10)\end{tabular} \hspace{-0.8em}
& \hspace{-0.6em}\begin{tabular}[c]{@{}l@{}}Input(24, 24, 3)\\ Conv1(5,5,1,16)\\ Pool1(2,2)\\ $	\overline{\mbox{Res/Conv1(5,5,16,32)}}$\\Res/Conv2(5,5,32,32)\\ \underline{Res/Conv3(5,5,32,32)}\\ Pool2(2,2)\\ Conv3(3,3,32,64)\\ Pool3(2,2)\\ FC1(576, 256)\\ FC2(256, 128)\\ FC3(128, 10)\end{tabular} \hspace{-0.8em}

\\ \hline
\end{tabular}
\caption{Summary of Applications.}
\label{table:applications}
\end{table}

With these requirements, we implemented the Shenjing components in Java. 
The partial-sum and spike NoCs of the functional simulator are essentially the corresponding RTL modules with all the details of registers, wirings, crossbars etc. rewritten in Java.
We verified this functional simulator against our RTL simulator, automatically checking the state of each component cycle by cycle given the same input instructions and data.  We have also verified our power estimation against the RTL simulation using the MNIST-MLP benchmark.

We simulate Shenjing with MNIST and CIFAR-10 datasets on 4 neural network structures, covering the most common neural network layers including convolution, average pooling, fully connected and residual block. 



\begin{table}[tb]
\centering
\footnotesize
\vspace{-1em}
\begin{tabular}{|c|c|c|c|c|}
\hline
         & MNIST  & MNIST  & CIFAR-10 & CIFAR-10 \\ 
         & MLP    & CNN    & CNN     & ResNet  \\ \hline
ANN Accu. & 0.9967 & 0.9913 & 0.7992  & 0.7825  \\ \hline
Abstract SNN Accu.  & 0.9611 & 0.9715 & 0.7590  & 0.7250 \\ \hline
Shenjing Accu. & 0.9611 & 0.9715 & 0.7590  & 0.7250  \\ \hline
\#Cores  & 10    & 705    & 2977    & 5863  \\ 
  &      &     & (4 chips)    & (8 chips)  \\ \hline
Timestep (T)      & 20     & 20    & 80      & 80      \\ \hline

Frames per sec      & 40     & 30    & 30      & 30      \\ \hline

Frequency& 120~kHz & 207~kHz     & 1.25~MHz    & 2.83~MHz    \\ \hline
Power (mW)& 1.35     & 87.54   & 456.71    & 887.81   \\ \hline
Power/Core (mW)& 0.135     & 0.124   & 0.153    & 0.151   \\ \hline
mJ/frame& 0.038   & 2.92      & 15.22    & 29.59  \\ \hline
Mapping time(ms)& 660   & 2142      & 4384    & 12022  \\ \hline 
\end{tabular} 
*Timestep is the length of spike train for each image.
\caption{Overall Performance.}
\label{table:benchmark}
\end{table}

\noindent
{\bf Accuracy.} 
Table~\ref{table:benchmark} reports the accuracy of the original ANN model, the converted abstract SNN model, and the SNN model when implemented on Shenjing. 
SNN's use of binary instead of full precision floating point data brings about a loss in accuracy. This is the main disadvantage of SNNs. However, for most applications, the amount of accuracy loss is usually acceptable. In fact, there is substantial research on reducing the resource requirements of ANNs at the cost of accuracy loss.

MNIST: We built both a fully connected as well as a CNN network for MNIST as shown in Table~\ref{table:applications}(a-b). 
As MNIST is a relatively simple dataset, both networks achieved high accuracy when trained as ANN. The converted SNN suffers just 3\% loss in accuracy before mapping to hardware, with no additional loss after mapping onto Shenjing. 

CIFAR-10 CNN:
CIFAR-10 is a dataset of 60,000 $32\times32$ color images in 10 classes, 6000 each. The image size was first reduced to $24\times 24$ by center cropping. We built a conventional CNN and a residual CNN for it following the methods described in~\cite{Cao2015} as shown in Table~\ref{table:applications}(c). The resultant SNN achieved an accuracy of 75.90\%, while the ANN accuracy was 79.92\%.

CIFAR-10 ResNet: Residual network (ResNet)~\cite{he2016deep} is designed to solve the degradation problem in training very deep CNNs. In this benchmark, we demonstrate that Shenjing supports ResNet using a small scale network as in Table~\ref{table:applications}(d).
To the best of our knowledge, this is the first demonstration of a SNN hardware that can be configured automatically to run residual networks, thanks to Shenjing's NoCs.

\noindent
{\bf Mapping time.} Our toolchain automatically mapped all the above ANNs onto Shenjing. 
The mapping for the largest network took 12 seconds on an Intel Core i7-8550U CPU.

\noindent
{\bf Power.}
Table~\ref{table:benchmark} shows our architectural power estimates. 
Active power is estimated by multiplying the synthesized active energy numbers per atomic operation (Table~\ref{table:energy}) with the count of each atomic operation obtained from our functional simulator and dividing the sum by running time.
 For multi-chip applications (CIFAR-10 CNN and ResNet), we assume 4.4pJ/bit for inter-chip I/O, based on state-of-the-art 56Gbps serial link on 28nm (same process as Shenjing)~\cite{shibasaki20163}. 

MNIST MLP is estimated at 120KHz frequency for validation of our architectural power estimates against RTL power numbers. Our functional simulator estimated total power for this network to be at 1.35mW, which is close to the detailed RTL-based power estimate of  1.26mW. The other models are simulated at different frequencies, targeting 30fps throughput.

%
%
\section{Related works}
\label{sec:related-works}
%

\begin{table}[tb]
\footnotesize
\vspace{-1em}

\begin{tabular}{|c|c|c|c|c|c|c|}
\hline
 Architecture                        & Tech            & Accu. & FPS                   & Voltage     &Power& uJ/\\

& (nm)&   & & &(mW) &Frame\\ \hline

 SNNwt~\cite{du2015neuromorphic}                       & 65                  & 91.82\%  & N.A.                    & 1.2V        &N.A.& 214.7                                                                   \\ \cline{1-7}

    SpiNNaker                       & 130                  & 95.01\%  & 77                    & 1.8V/        &300& 3896     \\ 
    ~\cite{spinnaker}& &   & &1.2V & & \\ 
\cline{1-7} 

 Tianji~\cite{ji2018bridge}                       
  & 120                 & 96.59\%  & N.A.                    & 1.2V        &120*& N.A.                                                          \\ \cline{1-7} 

 TrueNorth   & \multirow{2}{*}{28} & 92.70\%  & \multirow{2}{*}{1000} & \multirow{2}{*}{0.775V}      & 0.268 &0.268                                                                 \\ \cline{3-3} \cline{6-7} 
 ~\cite{esser2015backpropagation} &                       & 99.42\%  &           &                     & 108 &108                                                                   \\ \cline{1-7} 
 {\it This work }                      & {\it 28}                  & {\it 96.11\%}  & {\it 40}                    & {\it 1.05V/} & {\it 1.26} & {\it 38}                                                                    \\ 
     & &   & & {\it 0.85V} & & \\ 
\hline
\end{tabular}
* is only dynamic power.
\caption{Comparison with existing SNN architectures for MNIST MLP}
\label{table:hardware-comparison}
\end{table}

The field is huge so we can only focus on SNN accelerators and NoCs in neural network chips here.

\noindent
{\bf SNN Hardware.} 
Apples-to-apples comparison with prior SNN architectures is difficult especially when we do not have access to others' models and implementations.
Table~\ref{table:hardware-comparison} is our best-effort comparison with existing SNN architectures for MNIST benchmark. 
The spatially expanded SNNwt~\cite{du2015neuromorphic} is an application specific architecture where input size of neurons are precisely 784, hence avoiding the partial-sum issue, but it will not scale. SpiNNaker~\cite{spinnaker} uses 20 ARM cores to implement neurons and facilitates communication with two specialized NoCs. One NoC handles inter-chip communication in a 2D triangular toroidal mesh and the other handles communication between cores and peripherals.
Loihi~\cite{loihi}, like SpiNNaker, uses conventional CPUs and hence incurs high power. 

IBM's TrueNorth fabricated 4096 cores connected by a mesh NoC~\cite{truenorth}. Due to the rigid 256$\times$256 input/output constraints of each core, migrating any neural net to TrueNorth requires much restructuring and retraining of the model, often ending up using more resources. In fact, TrueNorth has its own programming paradigm and simulator~\cite{amir2013cognitive} that users have to program, train and optimize for. Energy is highly correlated with accuracy in TrueNorth: its power increases by 402$\times$ when the MNIST accuracy is boosted from 92.7\% to 99.42\% (from Figure 3C of~\cite{esser2015backpropagation}) because a different model had to be used. Shenjing achieves an accuracy of 96.11\% when running MNIST, with energy that is an order of magnitude lower than SNNwt~\cite{du2015neuromorphic}, comparable to TrueNorth. 
For CIFAR-10, with significant manual tuning, TrueNorth produces 83.41\% accuracy with 4042 cores at 204.4 mW~\cite{Esser11441}. In comparison, we simply automatically map an existing neural net using 2977 cores and achieved a 75.9\% accuracy close to that of the original ANN version. Shenjing consumes a higher 664.20 mW, a large portion of that (47\%) is consumed by the SRAMs. 
Implementation wise, TrueNorth adopts custom SRAMs and circuit layouts at lower voltage and mixed asynchronous-synchronous circuits, while Shenjing uses the foundry's memory-compiled higher-VDD SRAMs and fully synthesized clocked circuits.

Ji et. al.~\cite{ji2018bridge} proposed a software mapping tool that takes ANN/SNN models and automatically transforms and maps them onto hardware accelerators taking into account hardware constraints. They demonstrated mapping of ANNs onto the Tianji SNN accelerator~\cite{shi2015development}, enabling transfer learning of trained ANN models. However, without partial sum support in the NoCs, they have to transform the neural nets substantially for good accuracy. Shenjing's NoCs enable the hardware to realize same accuracy as the original abstract SNN model, simplifying mapping. The recent Tianjic chip~\cite{pei2019towards}, which we believe is the further development of~\cite{shi2015development}, configures cores into ANN, SNN or hybrid mode. It handles sums across cores by configuring some cores into hybrid mode and adding an extra ANN layer for accumulation. Compared to Shenjing's PS NoCs, this approach lowers core utilization for computing and requires restructuring of SNN models. The 156-core Tianjic chip consumes 0.95W.



\noindent

\noindent
{\bf Related NoCs.} Prior NoCs in SNNs for delivering spikes are routed dynamically. Hence, spike trains may not be delivered in global order to all neuron cores, causing errors in spiking. Prior works tackle this with a fixed-latency interconnect~\cite{embracepc2013} or by a hardware ordering mechanism within the NoC~\cite{oridac2017}. 
Shenjing's spike NoC is software-scheduled, and the scheduling algorithm ensures ordering of the spikes so the routers do not need to handle ordering.
Computation in NoCs has been proposed in~\cite{kwon2018maeri}, configuring adder switches at run-time to support arbitrary sizes for DNN partition and mapping whilst our adders in partial sum NoCs are fixed. 

%
%
%
%
\section{Conclusion}
\label{sec:conclusion}
In this paper we introduced Shenjing, a SRAM-based SNN accelerator for on-device AI. Unlike previous works on SNNs that attempt to mimic the brain, Shenjing has a more practical goal of providing an architectural means to map trained ANNs onto energy-efficient SNNs. The key to enable such transfer learning in a scalable way lies in the software-defined partial sum NoCs and per-neuron spike NoCs. This paper also outlined the mapping algorithm that will exploit these NoCs efficiently when mapping ANNs into SNNs. Our results show that on a 28nm technology node, Shenjing is able to implement the CIFAR-10 ResNet deep learning benchmark using 5863 cores and  0.887W while achieving 72.50\% accuracy. 

\bibliographystyle{ieeetr}

\bibliography{ref}

\end{document}